\documentclass[pdflatex,sn-nature]{sn-jnl}

\usepackage{graphicx}%
\usepackage{multirow}%
\usepackage{amsmath,amssymb,amsfonts}%
\usepackage{amsthm}%
\usepackage{mathrsfs}%
\usepackage[title]{appendix}%
\usepackage{xcolor}%
\usepackage{textcomp}%
\usepackage{manyfoot}%
\usepackage{booktabs}%
\usepackage{algorithm}%
\usepackage{algorithmicx}%
\usepackage{algpseudocode}%
\usepackage{listings}%

\theoremstyle{thmstyleone}%
\theoremstyle{thmstyletwo}%

\theoremstyle{thmstylethree}%

\begin{document}

\title[Improving global precipitation forecasts with an AI weather model trained on satellite observations]{\vspace{-2cm}Improving global precipitation forecasts with an AI weather model trained on satellite observations}

\author[1]{\fnm{Julian F.} \sur{Schmitt}}\email{julianschmitt@google.com}

\author[1]{\fnm{Bertrand} \sur{Delorme}}\email{bdelorme@google.com}

\author[1]{\fnm{Robert C.} \sur{King}}\email{robcking@google.com}

\author[2]{\fnm{Yashica} \sur{Patodia}}\email{yashicap@google.com}

\author[2,3]{\fnm{Tapio} \sur{Schneider}}\email{tapio@caltech.edu}

\author[4]{\fnm{Aditi} \sur{Sheshadri}}\email{Aditi\_Sheshadri@stanford.edu}

\author*[1]{\fnm{Ravi} \sur{Jain}}\email{ravijain@google.com}

\affil[1]{\orgdiv{X, the Moonshot Factory}, \orgname{Google LLC}, \orgaddress{\street{100 Mayfield Ave}, \city{Mountain View}, \postcode{94043}, \state{CA}, \country{USA}}}

\affil[2]{\orgdiv{Google}, \orgname{Google LLC}, \orgaddress{\street{1600 Amphitheatre Pkwy}, \city{Mountain View}, \postcode{94043}, \state{CA}, \country{USA}}}

\affil[3]{\orgdiv{Environmental Science and Engineering}, \orgname{California Institute of Technology}, \orgaddress{\street{1200 E California Blvd}, \city{Pasadena}, \postcode{91125}, \state{CA}, \country{USA}}}

\affil[4]{\orgdiv{Earth System Science}, \orgname{Stanford University}, \orgaddress{\street{450 Jane Stanford Way}, \city{Stanford}, \postcode{94305}, \state{CA}, \country{USA}}}


\abstract{Precipitation forecasts shape decision-making across the global economy, particularly in sectors such as agriculture. However, unlike variables such as temperature, precipitation is highly intermittent and localized, making it difficult to forecast. While recent advances in AI weather prediction systems have enabled them to surpass physical models on globally averaged metrics, improvements in mean error rarely translate to actionable forecasts of severe flooding or dry crop fields. Furthermore, most of these models are trained and evaluated against a reanalysis data product, ERA5, which has well-known biases. Here we retrain AIFS, ECMWF’s widely-used, open-source operational 0.25$^\circ$ probabilistic graph-transformer weather model, on satellite-based precipitation observations. Our model, Laxmi, improves global probabilistic accuracy by 19\% and resolves systematic distributional biases in ERA5. Specifically, Laxmi reduces drizzle overprediction by 33\% for amounts less than 3~mm~day$^{-1}$. It also mitigates extreme rainfall underprediction, improving the global 95\textsuperscript{th} percentile Brier skill score by 57\%. Across a case study of 10 Indian tropical storms, Laxmi delivered the most accurate forecast of 150~mm event-total precipitation in 7 events, compared to 1 for AIFS and 2 for the leading physical model, IFS. Our results demonstrate that incorporating observation-based precipitation data directly into training can substantially improve forecasts.}




\maketitle

\section*{Introduction}\label{sec1}

Precipitation forecasts shape decision making across the global economy and are particularly important for agriculture~\cite{Burlig2024-cj}. Despite recent advances, evaluating and predicting sparse, intermittent, and localized precipitation remains uniquely challenging: global observational datasets often disagree, and rain gauge data poorly represent grid-box averages~\cite{Rivoire2021-zw}. Simultaneously, weather models must approximate the processes that lead to precipitation, such as convection and microphysics, because they occur at scales too small to be explicitly resolved~\cite{Schneider2017-kb}. While recently developed Artificial Intelligence Weather Prediction (AIWP) models offer a significantly faster, cheaper, and often more accurate~\cite{Brenowitz2025-la} complement to traditional forecasting systems, current models consistently over-forecast drizzle and under-forecast extreme weather events~\cite{Zhang2026-vh, Bonev2025-we, Sun2025-nj, Qiang-Sun2025-zl}. Correcting these biases is a priority as AIWP models enter operational deployment~\cite{Masiwal2026-jy, Lang2026-pd}. 

These precipitation errors stem from two distinct challenges: the models' loss function and training data biases. Standard deterministic loss functions suffer from a double-penalty bias when precipitation is merely shifted in space; effectively, they express uncertainty as a smoothing of the forecast field, resulting in overly smooth forecasts that underestimate the highest intensity events and over-predict drizzle~\cite{Lam2023-tj}. While recent probabilistic models mitigate some of this smoothing~\cite{Price2025-yb}, their accuracy remains fundamentally constrained by biases in the training data itself. Most AIWP models are trained on ERA5 (ECMWF Reanalysis version 5)~\cite{Hersbach2020-lj} that relies on the same microphysical and convective parameterizations that make conventional precipitation forecasting challenging. By emulating the larger-scale effects of these parameterizations, AIWP models inherit their structural biases; for ERA5, this includes a systematic overprediction of drizzle~\cite{Pradhan2025-ng} and underprediction of high-intensity rainfall~\cite{Lavers2022-cj, Rivoire2021-zw, Henin2018-fi}. In operational settings, these biases can lead to a failure to issue warnings ahead of extreme events, or delayed or suspended harvests in operations such as sugarcane harvesting~\cite{Vieira2019-wo}. Satellite-derived observational datasets, such as IMERG (Integrated Multi-satellite Retrievals for GPM)~\cite{Huffman2015-xh} provide an alternative that does not inherit parameterization biases but carries its own retrieval and calibration biases. However, observational precipitation datasets are not everywhere consistent with ERA5 atmospheric variables, such as specific humidity and vertical velocity, which physically give rise to precipitation, making training at high resolution challenging with model architectures that respect such physical constraints~\cite{Yuval2026-dd}.

Here, we demonstrate that an AI model architecture, a graph-transformer, can be used to successfully map the ERA5 atmospheric state to high-resolution satellite observations to mitigate systemic drizzle and extreme-precipitation biases in current operational AIWP models. To do so, we retrain ECMWF's operational AI Forecasting System (AIFS-CRPS)~\cite{Lang2026-pd} directly on the IMERG dataset. To isolate the impact of the training data, we retain the same graph-transformer architecture, Continuous Ranked Probability Score (CRPS) loss function, and core ERA5 atmospheric variables as in AIFS-CRPS, but substitute IMERG data for the target precipitation field; we dub this model Laxmi. We find Laxmi improves systematic precipitation intensity biases and medium-range accuracy when compared to the operational AIFS-CRPS~\cite{Lang2026-pd} and ECMWF's physics-based ensemble Integrated Forecast System (IFS ENS, abbreviated to IFS henceforth)~\cite{European-Centre-for-Medium-Range-Weather-Forecasts2024-qt}. 

\section*{Results}
\subsection*{Improved precipitation distribution}
Selecting a target dataset for training an AIWP model requires balancing accuracy with spatial coverage and temporal resolution. While ground-based radar-gauge datasets, such as the U.S.\ Multi-Radar Multi-Sensor (MRMS) system~\cite{Smith2016-nm}, are considered closest to ground truth~\cite{Beck2019-sb}, they lack coverage across much of the tropics, precluding their use in global models. Reanalyses, such as ERA5~\cite{Hersbach2020-lj}, circumvent the lack of coverage by using a physical model to assimilate observations. ERA5 is widely considered to be the best reanalysis~\cite{Beck2019-sb}, yet the convection parameterization in the underlying physical model produces too frequent precipitation events, increasing the frequency of light rain and decreasing the frequency of heavy rain compared to gauge data~\cite{Su2025-rm}. Compared to several satellite products, ERA5 produces precipitation about twice as frequently in the tropics and at the wrong time of day~\cite{Pradhan2025-ng}, and it underestimates precipitation associated with tropical mesoscale convective systems---which account for over 50\% of tropical precipitation---by 25-34\%~\cite{Heflin2026-mg}. Over the Indian subcontinent, IMERG more accurately reproduces the observed precipitation intensity--temperature relation than ERA5 when validated against gauge-based measurements~\cite{Sengupta2023-un}.

\begin{figure}
    \centering
    \includegraphics[width=.7\textwidth]{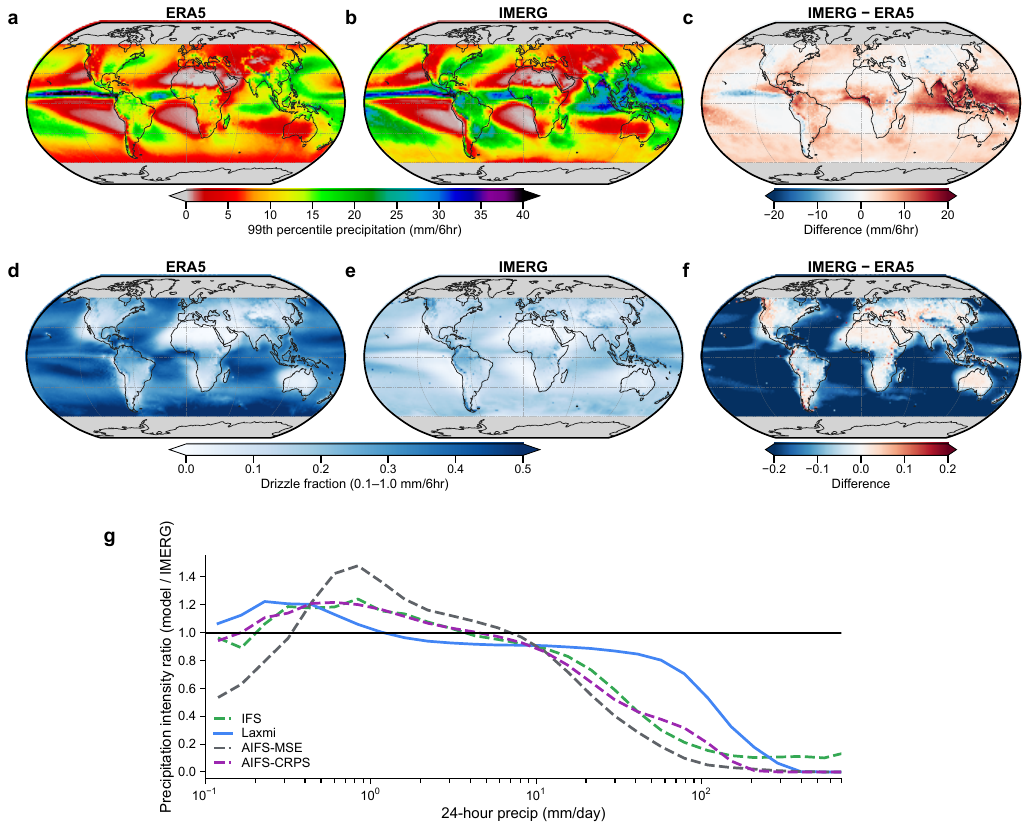}
    \caption{\textbf{Global precipitation intensity distributions compared to IMERG observations.} 99\textsuperscript{th} percentile precipitation for (\textbf{a}) ERA5, (\textbf{b}) IMERG, and (\textbf{c}) the difference computed between 2001 and 2023. Fraction of 6-hour time windows which experience between 0.1 and $1\;\mathrm{mm\;6 hr^{-1}}$ for (\textbf{d}) ERA5, (\textbf{e}) IMERG, and (\textbf{f}) the difference, computed between 2001 and 2023. Mean precipitation rates over the period are shown in Supplementary Fig.~1, and regional ERA5-to-IMERG frequency ratios are shown in Supplementary Fig.\~2. . (\textbf{g}) Precipitation intensity ratios, or the model frequency divided by observed IMERG frequency for a given precipitation intensity, are evaluated at 0.25$^\circ$ spatial resolution for IFS, Laxmi, AIFS-MSE, and AIFS-CRPS. To compute these ratios, we aggregate 24-hour accumulated precipitation for 94 forecasts for each model for all lead times up to 15 days. IMERG frequencies are computed over the global domain between 60$^\circ$S and 60$^\circ$N between January 2023 to September 2025. A ratio of 1, shown with a horizontal black line, indicates that a model perfectly emulates the observed precipitation frequency distribution of IMERG. Solid lines distinguish models trained with IMERG precipitation observations, whereas dashed lines denote the physics-based IFS and AIWP models trained on ERA5 precipitation. We exclude observations that are less than $0.1 \mathrm{\,mm \;day{^{-1}}}$ and truncate at the high end above $908 \mathrm{\,mm\; day^{-1}}$ where the baseline IMERG probability density drops below $10^{-10}$. The probability densities before a ratio is taken are shown in Supplementary Fig.~3.}
    \label{figure1}
\end{figure}

We select IMERG as our observational target, as it consistently ranks as the leading global satellite dataset when evaluated against radar-gauge networks~\cite{Beck2019-sb}. The structural differences between IMERG and ERA5 are most visible when comparing precipitation intensity and frequency (Fig.~\ref{figure1}), both of which are important for agricultural applications. At the 99\textsuperscript{th} percentile, ERA5 under-produces precipitation compared to IMERG (Fig.~\ref{figure1}\textbf{a}-\textbf{c}) by over a factor of two across the Maritime Continent, with notable deficits also occurring in major agricultural regions such as India, Brazil, West Africa, and the central and eastern U.S.

In addition to missing extremes, ERA5 exhibits a pervasive drizzle bias. While the bias is largest over the ocean, IMERG also exhibits notably less drizzle across the same or similar agricultural regions (Fig.~\ref{figure1}\textbf{d}-\textbf{f}). Accurately forecasting continuous dry periods in these regions is vital for agricultural operations, which often require continuous dry windows for planting and harvest~\cite{Vieira2019-wo}. By training directly on IMERG, Laxmi avoids inheriting these structural errors.

To quantify the learned distribution shift from training on IMERG, we evaluate the global precipitation intensity ratio (Fig.~\ref{figure1}\textbf{g}). Alongside IFS and AIFS-CRPS, we include the deterministic AIFS-MSE~\cite{Moldovan2025-fz} to highlight recent improvements in loss function development. We find Laxmi more accurately reproduces the IMERG precipitation distribution across the tropics and extratropics (Supplementary Fig.~3), reducing drizzle (${<}3$\,mm\,day$^{-1}$),  and better capturing the frequency of precipitation above $10~\mathrm{mm\;day^{-1}}$. While all models struggle at extreme intensities, Laxmi captures 68\% of the observed events exceeding 50~mm\,day$^{-1}$, compared to $\sim$30\% for both AIFS-CRPS and IFS. Above 100~mm\,day$^{-1}$, Laxmi captures 38\% of the observed events, which is double that of IFS (17\%) and nearly 10 times more than AIFS-MSE (4\%). Only at the extreme tail (${>}$300~mm\,day$^{-1}$) does Laxmi fail to produce precipitation, a domain where the physics-based IFS still captures about 10\% of observational frequency.

\subsection*{Global medium-range skill}
\begin{figure}
    \centering
    \includegraphics[width=\linewidth]{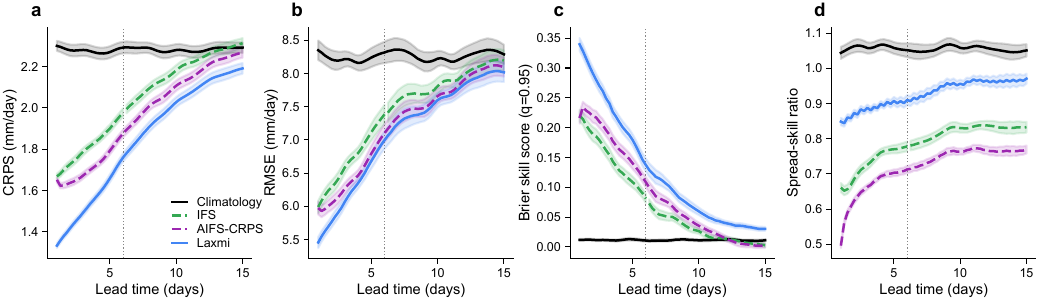}
    \caption{\textbf{Forecasting scores for 24-hour accumulated precipitation evaluated against IMERG.} Global evaluation metrics calculated between 60$^\circ$S and 60$^\circ$N at 0.25$^\circ$ resolution for (\textbf{a}) CRPS, (\textbf{b}) unbiased ensemble mean RMSE, (\textbf{c}) Brier skill score at the 95\textsuperscript{th} percentile, and (\textbf{d}) Spread-Skill Ratio. Solid curves indicate models trained on IMERG, including Laxmi and a baseline climatology; dashed curves represent models trained against ERA5 or physics-based baselines. Each model was initialized 94 times with 50 ensemble members weekly at 6:00 and 18:00 UTC from February 2024 to December 2024. The 6:00 and 18:00 UTC cycles were selected to prevent future-data leakage from ERA5's 12-hour assimilation windows~\cite{Lam2023-tj}. The vertical dotted line at Day 6 denotes the extent of IFS forecasts initialized on these cycles. Subsequent scores for IFS use the 00:00 and 12:00 UTC cycles. Climatology is generated with 50-member ensembles drawn from historical IMERG observations using the same dates as the models~\cite{Yuval2026-dd}. Each ensemble is sampled without replacement at the same time of day as the corresponding forecast, and within 7 days of a given initialization day of year. This sampling approach allows us to statistically account for changes in climatology over the course of the year; climatological error at any given lead time is the average over these samples. Climatological error therefore retains some variability as a function of lead time. Shading for each curve represents a 95\% confidence interval for the average score over the initialization times. An evaluation at 1$^\circ$, including additional models trained at 1$^\circ$, is shown in Supplementary Fig.~4. Systematic evaluations of Brier scores and Brier skill scores at the 90\textsuperscript{th}, 95\textsuperscript{th}, and 99\textsuperscript{th} precipitation percentiles are shown in Supplementary Figs.~5 and~6. Model evaluations restricted to the tropics are provided in Supplementary Figs.~7--9.}
    \label{figure2}
\end{figure}

We next evaluate Laxmi's medium-range performance across 94 ensemble forecasts over the 2024 holdout year using WeatherBenchX~\cite{Rasp2024-zq, Agrawal2025-pv} against global ($60^\circ\text{S}$--$60^\circ\text{N}$) IMERG observations. Evaluated at $0.25^\circ$ spatial resolution, Laxmi demonstrates substantial improvements over both operational AIFS-CRPS and IFS (Fig.~\ref{figure2}).

At a 1-day lead time, Laxmi reduces 24-hour accumulated precipitation CRPS by 19.3\% compared to AIFS-CRPS and 20.1\% compared to IFS. Unbiased ensemble mean root-mean-square error (RMSE) drops by 8.7\% and 8.9\%, respectively. The performance gap widens for heavy precipitation: at the same 1-day lead time, Laxmi improves the Brier Skill Score at the 95\textsuperscript{th} percentile by 57.2\% over AIFS-CRPS and 54.4\% over IFS. For all metrics, Laxmi is more skillful than both models across the entire 15-day forecast horizon.

In addition to improving global error metrics, fine-tuning on IMERG also improves ensemble calibration. Overconfidence, or an underspread ensemble, is a common challenge in AIWP models~\cite{Price2025-yb} as optimization tends to reduce variance for localized events like precipitation~\cite{Moldovan2025-fz}. Although probabilistic loss functions, such as CRPS, yield substantial improvement~\cite{Lang2026-pd}, reanalyses still underestimate precipitation variance due to parameterization shortfalls in the underlying model~\cite{Sun2006-bm}. We quantify calibration via the spread-skill ratio, where 1 denotes a perfectly calibrated ensemble that is neither overconfident (ratio ${<}1$) nor underconfident (ratio ${>}1$). Fine-tuning directly on IMERG substantially reduces overconfidence: Laxmi achieves a ratio of 0.83 at a 1-day lead time and exceeds 0.97 by day 15. In contrast, AIFS-CRPS and IFS have ratios of 0.50 and 0.66 at day 1 and never exceed 0.77 and 0.83, respectively. By reducing overconfidence, Laxmi provides a more trustworthy assessment of its own forecast probabilities.

Similar to Lang et al.~\cite{Lang2026-pd}, we find a substantial improvement on medium-range metrics when using a higher-resolution model. We compared Laxmi against two coarser models trained on IMERG precipitation: a 1$^\circ$ graph-transformer, IMERG-O96-CRPS (Supplementary Fig.~4), and NeuralGCM-precip at 2.8$^\circ$~\cite{Yuval2026-dd} (Supplementary Fig.~10). When conservatively regridded to each model's native resolution, Laxmi outperforms both across all error metrics. The sole exception is the spread-skill ratio, where all three models perform similarly. Rather than a uniform increase in skill with resolution, we find different error growth profiles when Laxmi is compared to each of the lower-resolution models. When compared to NeuralGCM-precip at 2.8$^\circ$, we find that the largest skill gains are at the shortest lead times, consistent with better emulation of small-scale processes and higher topographic detail. However, when compared to IMERG-O96-CRPS at 1$^\circ$, the largest benefit occurs at 5- to 12-day lead times. This contrast suggests that, unlike for physics-based systems, the observed performance gains cannot be attributed to spatial resolution alone; the architecture and training schedule also dictate how errors propagate at longer lead times. 

\subsection*{A case study of Indian tropical storms}
\begin{figure}
    \centering
    \includegraphics[width=\textwidth]{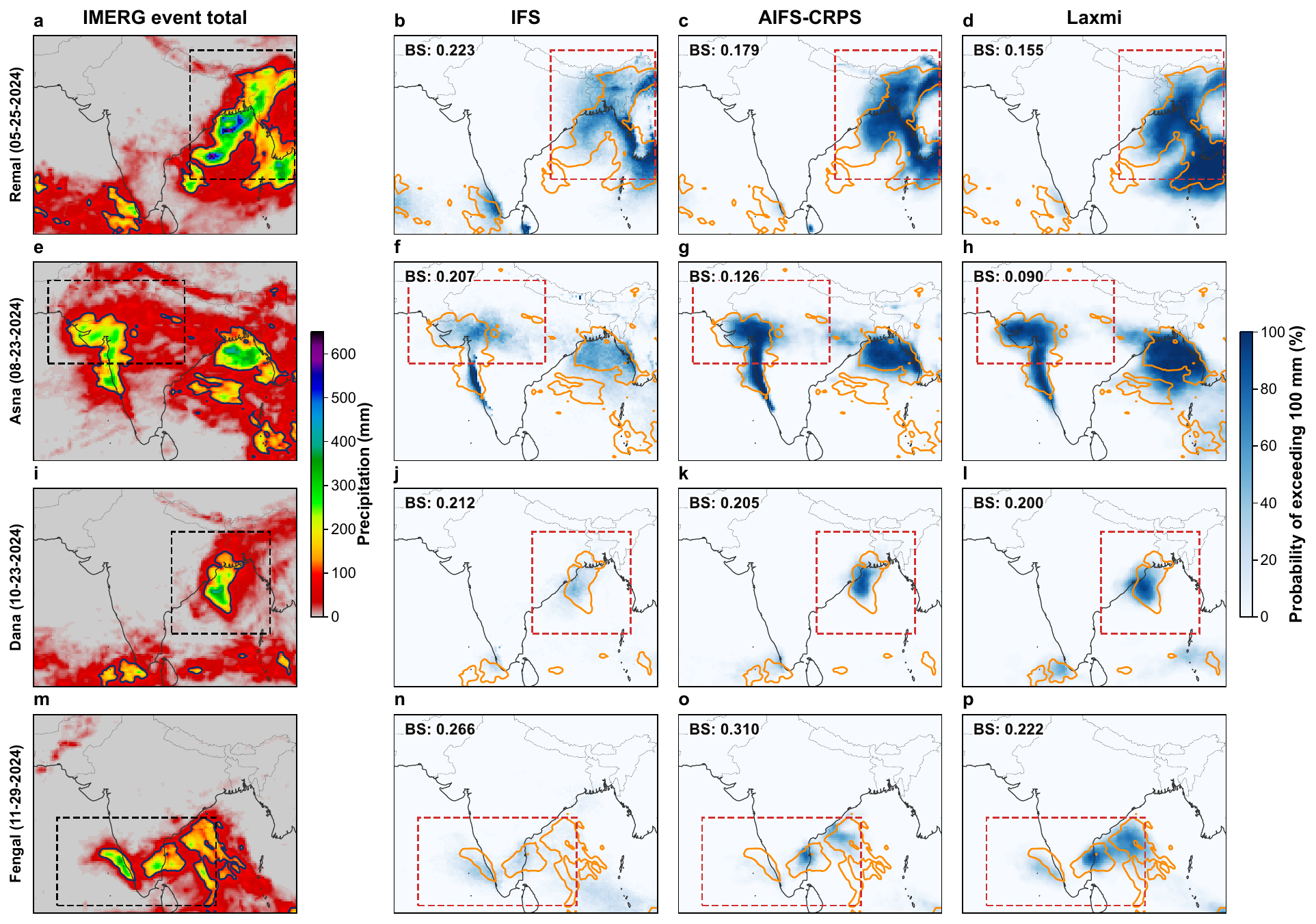}
    \caption{\textbf{Observed and forecasted extreme precipitation for four landfalling tropical storms in India.} Rows display event-integrated accumulation (spanning 1 day before to 3 days after landfall) and landfall date for storms Remal (\textbf{a}--\textbf{d}), Asna (\textbf{e}--\textbf{h}), Dana (\textbf{i}--\textbf{l}), and Fengal (\textbf{m}--\textbf{p}). Columns show observed IMERG accumulated precipitation (\textbf{a}, \textbf{e}, \textbf{i}, \textbf{m}) alongside model probability of exceedance (PoE) over 50 member ensembles at the 100~mm threshold for IFS (\textbf{b}, \textbf{f}, \textbf{j}, \textbf{n}), AIFS-CRPS (\textbf{c}, \textbf{g}, \textbf{k}, \textbf{o}), and Laxmi (\textbf{d}, \textbf{h}, \textbf{l}, \textbf{p}) Annotated values in each panel report the event-level Brier score evaluated at the 100~mm accumulation threshold defined across the active region (see Methods), which is indicated by the dashed box.}
    \label{figure3}
\end{figure}

To evaluate real-world early-warning capability, we examine hindcasts of the tropical monsoon systems making landfall in India over the evaluation period, a region chosen due to its heavy reliance on precipitation for agriculture and well-known vulnerability to extreme rainfall~\cite{Bearpark2025-av}. Forecasts of precipitation totals for four named storms (Asna, Dana, Fengal, and Remal) are shown in Figure~\ref{figure3}~(see Methods). We evaluate Laxmi against both operational AIFS-CRPS~\cite{Lang2026-pd} and IFS~\cite{European-Centre-for-Medium-Range-Weather-Forecasts2024-qt}. Across the forecasts, we find our model more reliably places heavy precipitation in the correct locations.

\begin{figure}
    \centering
        \includegraphics[width=0.8\textwidth]{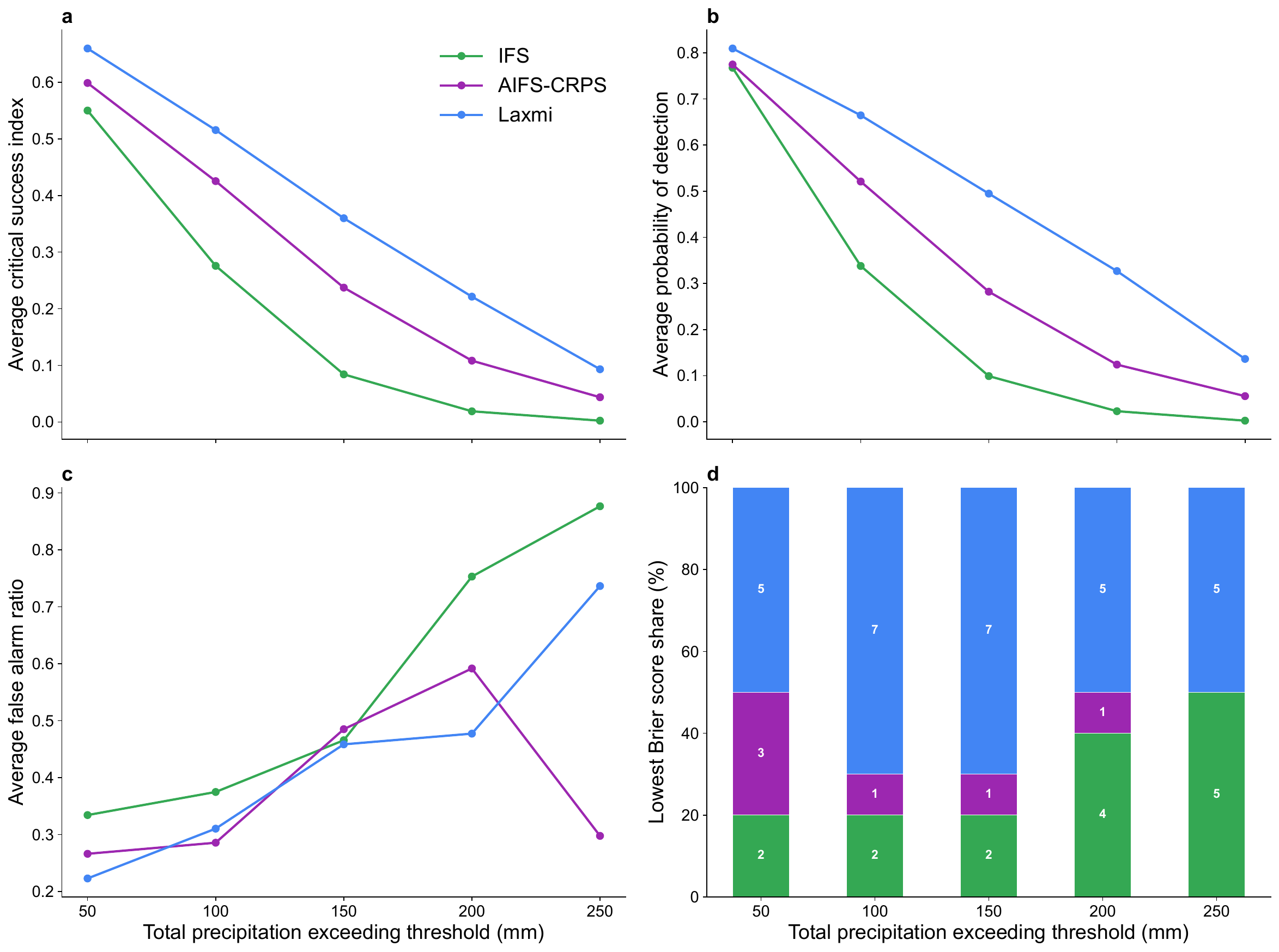}
        \caption{\textbf{Laxmi outperforms operational weather models on decision-oriented metrics.} Comparison of IFS, AIFS-CRPS, and Laxmi for 10 storms that reached depression strength or higher and made landfall in India between March 2024 and September 2025~\cite{Knapp2010-sx, R2019-cs}, evaluated across five precipitation thresholds (50, 100, 150, 200, and $250~\mathrm{mm\; day^{-1}}$). Panels show (\textbf{a}) Critical Success Index (CSI), (\textbf{b}) Probability Of event Detection (POD), (\textbf{c}) False Alarm Rate (FAR), and (\textbf{d}) frequency each model produced the lowest Brier score. Average Brier score values are shown in Supplementary Figure~11. Metrics are calculated over an active region, defined as the area within 500~km of the IBTrACS best track estimate, and further restricted to the union of the regions where any of the three models produce a non-zero probability of exceeding the precipitation threshold defined in Methods. Categorical metrics (CSI, POD, and FAR) are computed using a 30\% probability warning threshold (i.e., a grid cell is classified as a warning if at least 15 of 50 ensemble members exceed the precipitation target).}
    \label{figure4}
\end{figure}

Across all ten events, Laxmi achieves a competitively-low average False Alarm Rate (FAR) while producing the highest Probability of Detection (POD) and Critical Success Index (CSI), a metric that both rewards correct detection and penalizes false alarms (Fig.~\ref{figure4}\textbf{a}-\textbf{c}). This combination indicates that Laxmi is substantially better calibrated because it accurately forecasts heavy precipitation without additional false alarms. While IFS tends to overforecast heavy precipitation, as evidenced by its large FAR at high precipitation thresholds, Laxmi successfully detects heavy precipitation (high POD) without additional false alarms, suggesting that the model avoids a simple distributional shift toward more precipitation and instead captures both event location and intensity more accurately.

In addition to these categorical metrics, we also evaluate models on Brier score, which measures the accuracy of probabilistic predictions (calculated as the mean squared difference between the predicted probabilities and the observed outcomes). Across the 10 Indian tropical storms, Laxmi demonstrates the lowest average Brier score (Supplementary Fig.~11) across all precipitation intensity thresholds. However, evaluating how frequently each model delivers the best ensemble forecast, as measured by the lowest Brier score, demonstrates how Laxmi and IFS can complement each other operationally (Fig.~\ref{figure4}\textbf{d}). At the 50 and $100~\mathrm{mm}$ thresholds, Laxmi produces the best Brier score 12 of 20 times compared to 4 for AIFS-CRPS and 4 for IFS. However, at the $250~\mathrm{mm}$ threshold, IFS produces the best forecast 5 of 10 times with Laxmi producing the best forecast the remaining 5 times. AIFS-CRPS is not competitive at any threshold. Thus, while Laxmi is substantially better calibrated to observation-based precipitation than AIFS-CRPS, IFS forecasts are essential at very high precipitation thresholds.

As an additional demonstration of generalizability, we also evaluate the five hurricanes that made landfall in the United States over the evaluation period and find that Laxmi provides the best forecast, as measured by Brier score at the $100~\mathrm{mm\;day^{-1}}$ threshold, for four of the five events. The case studies are shown in Supplementary Figure~12 and the metrics for all 46 tropical cyclones globally are shown in Supplementary Figure~13.

\section*{Discussion}
Laxmi demonstrates the advantages of incorporating satellite-based observations into AIWP models. First, it achieves superior medium-range skill across global error metrics and produces a better-calibrated ensemble, yielding more accurate uncertainty quantification. Notably, it reduces CRPS by 19.3\% at day 1 and 3.4\% at day 10 compared to AIFS-CRPS, and rivals or exceeds the 6.4\% maximum improvement achieved by AIFS-CRPS over the physics-based IFS. Second, Laxmi addresses structural intensity biases associated with the standard AIWP training dataset: dramatically improving the representation of heavy precipitation (30 to $200~\mathrm{mm\;day{^{-1}}}$), and simultaneously reducing the pervasive overprediction of light rain~\cite{Pradhan2025-ng}. Third, for tropical storms, the model generates higher detection rates without inflating false alarms for moderate-to-heavy events. However, we find IFS, a physics-based model, remains essential for forecasting the highest precipitation events. Finally, we demonstrate that higher spatial resolution is crucial for AIWP precipitation accuracy; Laxmi (${\sim} 25\mathrm{\;km}$) systematically outperforms its ${\sim} 100\mathrm{\;km}$ variant and surpasses NeuralGCM-precip~\cite{Yuval2026-dd} (${\sim}280\;\mathrm{km}$) by over 10\%, while resolving features at ten times finer scale.

Further development should address the remaining intensity biases and explore training to other precipitation datasets. Progress will likely require loss functions that more aggressively penalize drizzle~\cite{Chen2024-ck}, alongside methods that resolve the spatial mismatch between satellite precipitation and the reanalysis atmospheric state to better capture heavy rainfall. Although training on IMERG successfully mitigates the drizzle and extreme-precipitation biases inherited from reanalyses, it introduces its own retrieval and calibration biases. Future efforts could fine-tune models on regional datasets, such as MRMS~\cite{Smith2016-nm}, the Indian Monsoon Data Assimilation and Analysis (IMDAA)~\cite{Rani2021-cc}, and networks of rain gauges. While each of these datasets has its own limitations, they provide superior targets for customizing forecasts to region-specific tasks. Finally, real-time deployment will require fine-tuning on operational analysis to ensure compatibility with real-time initial conditions. 

Our results indicate that further progress in AIWP is likely to be made by adding more satellite-based observations as fields into these models. Other currently available satellite-derived datasets, including ocean temperatures, soil moisture~\cite{Entekhabi2010-gi}, and radiation variables~\cite{Knapp2011-ha} could enhance the learned dynamics of AIWP models, thereby extending predictability horizons even further. 

\section*{Methods}\label{methods}

\subsection*{Data and pre-processing}
We rely on two primary datasets for model training and evaluation: the ECMWF Reanalysis version 5 (ERA5)~\cite{Hersbach2020-lj} and the Integrated Multi-satellitE Retrievals for GPM Version 07 Final (IMERG)~\cite{Huffman2015-xh}. Model variables match the operational AIFS configuration specified by Moldovan et al.~\cite{Moldovan2025-fz} and comprise a 13-level ERA5 atmospheric state and 31 surface variables. To prevent physical inconsistencies when substituting the target precipitation field with IMERG observations, we explicitly exclude ERA5 convective precipitation, surface runoff, and snowfall. The forward map is formulated as $x_{i+1} = \mathcal{M}(x_i, x_{i-1})$, where $x_i$ defines the discrete global atmospheric state at a 6-hour interval $i$.  

As IMERG observations (available from June 2000 onward) are distributed as instantaneous precipitation rates at a 30-minute, $0.1^\circ \times 0.1^\circ$ resolution, we convert them to 6-hour accumulated totals by multiplying the instantaneous rate by the duration and summing. For training, these accumulations are conservatively regridded to the model grids (O96, $\sim 1^\circ$; and N320, $\sim 0.25^\circ$). The training period for IMERG-trained models is June 2000 through December 2023, with 2024 and 2025 held out for validation. Global evaluations are strictly bounded between $60^\circ\text{S}$ and $60^\circ\text{N}$ because beyond these latitudes, IMERG passive microwave retrievals are masked over frozen surfaces and cannot be reliably substituted by geostationary infrared estimates~\cite{Tan2019-od}.

\subsection*{Model architecture and optimization}
Laxmi utilizes a graph-transformer architecture identical to AIFS-CRPS~\cite{Lang2026-pd} and was trained using the open-source \texttt{Anemoi} framework~\cite{Prieto-Nemesio2025-fg}. Like other more recent AIWP models~\cite{Price2025-yb, Lang2026-pd, Yuval2026-dd}, we use a probabilistic loss function, the almost fair Continuous Ranked Probability Score ($\text{afCRPS}_\alpha$), which substantially reduces the over-smoothing of the precipitation field. Training requires an ensemble to compute probabilistic scores and is therefore several times slower than training a deterministic model. During training, the forecast distribution $\{x_i\}_{i=1}^M$ is verified against a target state vector $y$ of ERA5 reanalysis or IMERG observations via the objective function defined by:
\begin{equation}
    \mathrm{afCRPS}_\alpha := \frac{1}{2M(M-1)}\sum_{i=1}^M \sum_{j=1}^M \left( |x_i - y| + |x_j-y| - (1-\epsilon)|x_i - x_j| \right)
\end{equation}
where $M=4$ is the ensemble size, $\epsilon := \frac{1-\alpha}{M}$, and $\alpha$ is a model hyperparameter governing the tradeoff between deterministic loss and ensemble spread, which we set to $\alpha = 0.95$.

Optimization is performed using the Adam optimizer ($\beta_1 = 0.9$, $\beta_2 = 0.95$, weight decay = 0.1) and a cosine learning rate schedule across a three-stage autoregressive rollout. Differing slightly from AIFS-CRPS~\cite{Lang2026-pd}, stage 1 uses a 1-step rollout for 75,000 steps with a maximum learning rate of $1.6\times 10^{-3}$. Stage 2 extends this to a 2-step rollout for 20,000 steps (maximum learning rate $1 \times 10^{-5}$). Stage 3 increments the rollout from 3 to 12 steps once per epoch for 3,100 steps (maximum learning rate $2\times 10^{-6}$). The batch size is 32 for N320 models and 20 for O96 models. Unlike previous implementations that utilized operational analyses for the final stage~\cite{Lang2026-pd}, we perform all three training stages against ERA5 due to data availability, except for precipitation, which is scored against IMERG. We trained the O96 models on 10 NVIDIA A100 (80 GB) GPUs and the N320 model on 128 NVIDIA A100 (40 GB) GPUs. Both configurations took approximately 10 days to train.

\subsection*{Probabilistic climatology baseline}
To establish a probabilistic baseline, we generate $M=50$ member climatological forecast ensembles by resampling historical IMERG observations~\cite{Yuval2026-dd}. For any given target initialization time $t$, we construct an ensemble of forecasts by resampling historical IMERG observations, $X_\text{hist}$. Each ensemble member, $m\in \{1, \dots, M\}$ is derived by selecting a random source time, $s_m$, where the year is drawn uniformly from 2001–2019 and the day of the year is selected within a 15-day window centered on the target date. While the source year and day vary, the specific time of day is held constant to match the target initialization. The resulting forecast for any lead time $\tau$ is defined by  $x_{\mathrm{clim}, m} = X_\text{hist} (s_m+\tau)$, with all $M=50$ members sampled without replacement. 

\subsection*{Indian monsoon case study setup}
To evaluate model performance during localized high-impact events, we analyzed hindcasts for all 10 tropical monsoon systems that made landfall in India between February 8\textsuperscript{th}, 2024 and September 30\textsuperscript{th}, 2025 and achieved depression strength or higher in the International Best Track Archive for Climate Stewardship (IBTrACS)~\cite{Knapp2010-sx, R2019-cs}. The evaluation period was bounded by the date when IFS began providing operational forecasts at 0.25$^\circ$ and the end of availability for our observational target, the IMERG v7 Final dataset. Forecasts for each event were initialized using the latest 06:00 or 18:00 UTC cycle that provided a lead time of at least 72 hours prior to the IBTrACS estimated landfall. Precipitation accumulation was calculated over a 4-day event-integrated window, spanning 1 day prior to landfall through 3 days post-landfall.

\subsection*{Evaluation metrics}\label{sec:eval_metrics}
Standard global medium-range forecast metrics, including CRPS, unbiased ensemble mean RMSE, and Spread-Skill Ratio are calculated using WeatherBenchX~\cite{Rasp2024-zq, Agrawal2025-pv}. 

To evaluate probabilistic predictions of extreme precipitation at a given threshold $q$ (e.g., the 95\textsuperscript{th} percentile), we convert an $M$-member ensemble forecast $\{x_m\}_{m=1}^M$ into an exceedance probability $p = \frac{1}{M}\sum_{m=1}^M \mathbb{I}(x_m > q)$, where $\mathbb{I}(\cdot)$ is the indicator function. We then compute the fair Brier score (abbreviated as Brier score or $\text{BS}$ throughout the text), which corrects for the variance penalty inherent in finite ensemble sizes ($M$):
\begin{equation}
    \text{BS}_{\text{fair}} = (p - o)^2 - \frac{p(1-p)}{M-1}
\end{equation}
where $o = \mathbb{I}(y > q) \in \{0, 1\}$ is the binary observation derived from IMERG precipitation $y$, and the second term applies the finite-sample correction. 

To measure skill relative to climatology, we calculate the Brier Skill Score (BSS) as:
\begin{equation}
    \text{BSS} = 1 - \frac{\text{BS}_{\text{fair}}}{\text{BS}_{\text{climatology}}}
\end{equation}
where $\text{BS}_{\text{climatology}}$ is the fair Brier score computed identically using the exceedance probability $p_{\mathrm{clim}} = \frac{1}{M} \sum_{m=1}^M \mathbb{I}(x_{\mathrm{clim}, m} > q)$ over the $M=50$ member resampled IMERG climatology. 

For localized event evaluations, such as evaluating heavy precipitation associated with storms, we constrain our calculations to an active region. The active region is defined as a bounding box covering the IBTrACS best-track estimate~\cite{Knapp2010-sx} for the 4-day accumulation period plus a $500\mathrm{\;km}$ buffer that is further restricted to grid cells where either the observations or any evaluated model outputs a non-zero probability of exceeding the target precipitation threshold.

To assess models within an operational decision-making context, we convert each forecast ensemble into a deterministic binary warning using a fixed threshold. The optimal choice of threshold, defined as a fraction of the ensemble which detects an event, varies based on the costliness of the event and the cost to take action. For events where the cost of inaction is the same as the cost of action, the best threshold is 50\%. Since large event-total precipitation generally has higher costs for inaction we select a threshold of 30\%~\cite{James2025-se, Richardson2000-mj}. Grid points within the active region are classified into hits ($H$), misses ($M$), and false alarms ($F$). From these frequencies, we compute the Probability of Detection ($\text{POD} = \frac{H}{H + M}$), the False Alarm Ratio ($\text{FAR} = \frac{F}{H + F}$), and the Critical Success Index ($\text{CSI} = \frac{H}{H + M + F}$).

\backmatter

\bmhead{Supplementary information}
Supplementary information is available for this paper.

\bmhead{Acknowledgments}
We thank Ferran Alet, Aaron Bell, Janni Yuval, Timothy Lee, Aman Gupta, Alfred Piccioni, William Boos, and Pedram Hassanzadeh for their help. Prithvi Nambiar, Mohil Patel, and Josh DeAndria contributed to technical discussions and related applications. Rachel Payne contributed to overall program strategic alignment and relevance to downstream application partners and use cases. Gemini was used to edit Python code for model evaluation and plotting and for editing text and figure captions; we reviewed and verified all code and output Gemini generated. 

\bmhead{Data availability}
The ERA5 dataset was downloaded and is available from the Climate Data Store (\url{https://cds.climate.copernicus.eu/}) and the IMERG data was downloaded and is available from the NASA Global Precipitation Measurement website (\url{https://gpm.nasa.gov/data/imerg}). The IBTrACS tropical storm tracks were downloaded and are available from NOAA (\url{https://www.ncei.noaa.gov/products/international-best-track-archive}). IFS ensemble forecasts were downloaded in July 2026 using Herbie~\cite{Blaylock2026-vf} and are provided by ECMWF~\cite{European-Centre-for-Medium-Range-Weather-Forecasts2024-qt}. 
\bmhead{Code availability}
For training we used the open source \texttt{Anemoi} framework~\cite{Prieto-Nemesio2025-fg} with all evaluations conducted using WeatherBenchX~\cite{Rasp2024-zq, Agrawal2025-pv}. Graphics were generated using matplotlib~\cite{Hunter2007-jw} and cartopy~\cite{Elson2024-zm}. Code will be made available if required for peer review upon request. 
\bmhead{Author contribution}
Conceptualization: A.S., R.J., B.D. Methodology: J.S., A.S., R.J., B.D. Software: J.S., B.D., R.K. Y.P. Formal analysis: J.S. Writing—Original draft: J.S., Writing—Review and Editing: J.S., T.S., R.J., R.K., B.D., A.S. Supervision: R.J., A.S. Project administration: R.J. Visualization: J.S., R.K.
\bmhead{Funding}
J.S., B.D., R.K., Y.P., R.J., and T.S. are employees of Alphabet. 
\bmhead{Competing interests}
B.D., R.J., and Y.P. own Alphabet stock. Professor Aditi Sheshadri’s contribution to this publication was as a paid consultant and was not part of her Stanford University duties or responsibilities. The authors declare no other competing interests related to the paper.
\bmhead{Materials availability}
Not applicable.
\bmhead{Consent for publication}
Not applicable.
\bmhead{Ethics approval and consent to participate}
Not applicable.


\bibliography{bibliography}

\end{document}


\maketitle

\subsection*{Note: Model Variants and Nomenclature}
While the main text focuses on our N320-resolution graph-transformer model, Laxmi, we also trained two lower-resolution models to isolate the effects of model resolution from the data the precipitation attention head was trained against. These additional models are evaluated only in the Supplementary Figures. To clearly identify all model variants in the Supplementary Figures we use a triplet nomenclature to separate the models that identifies the precipitation training data (IMERG or ERA5), the model resolution (O96 or N320), and the loss function (CRPS or MSE). Thus, Laxmi is equivalent to IMERG-N320-CRPS. The additional lower-resolution models are ERA5-O96-CRPS and IMERG-O96-CRPS. AIFS-CRPS would be ``ERA5-N320-CRPS''; however we preserve the name given to it in Lang et al.~\cite{Lang2026-pd}.
\newpage
\begin{figure}[ht]
    \centering
    \includegraphics[width=\textwidth]{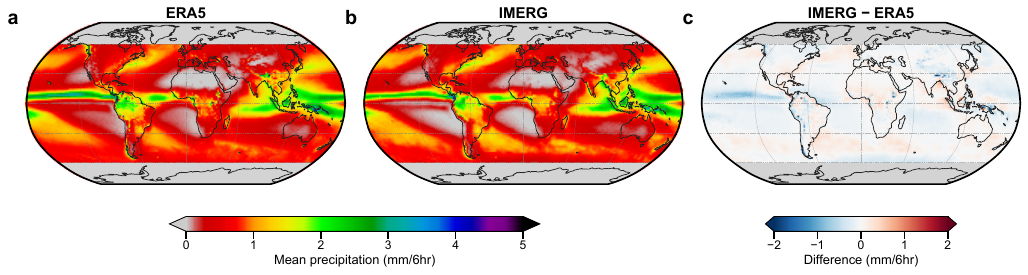}
    \caption{\textbf{ERA5 and IMERG mean precipitation climatology between 2000 and 2023 at 0.25$^\circ$.} (\textbf{a}) ERA5 mean precipitation, (\textbf{b}) IMERG mean precipitation, and (\textbf{c}) the difference in precipitation with red indicating that IMERG precipitates more at that location.}
    \label{fig:SI_mean_precip_imerg_era5}
\end{figure}

\begin{figure}[ht]
    \centering
    \includegraphics[width=\textwidth]{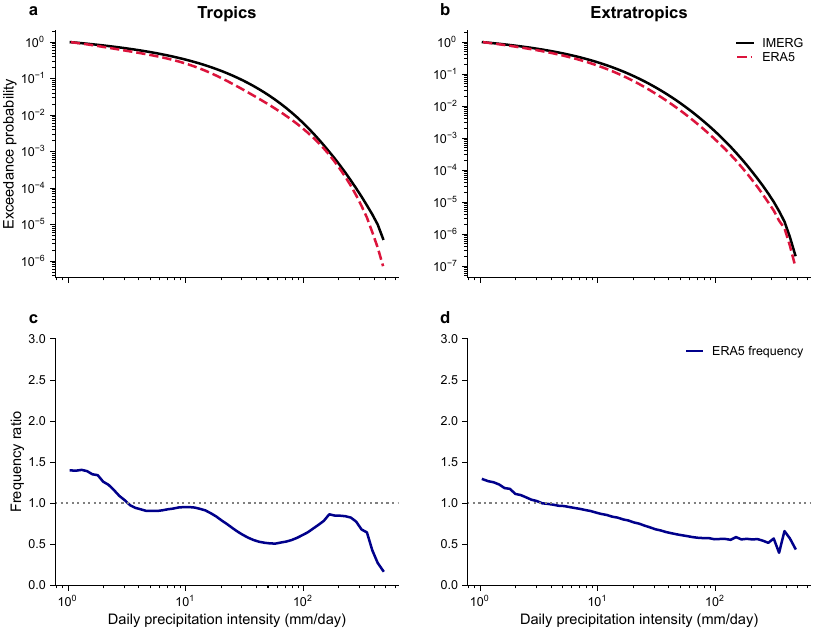}
    \caption{\textbf{Comparison of daily precipitation intensity distributions and frequency ratios between IMERG and ERA5.} Daily exceedance probability distributions at native N320 resolution are shown for (\textbf{a}) the tropics ($20^\circ\text{S}$--$20^\circ\text{N}$) and (\textbf{b}) the extratropics ($60^\circ\text{S}$ to $20^\circ\text{S}$ and $20^\circ\text{N}$ to $60^\circ\text{N}$). Corresponding relative frequency ratios ($\text{ERA5}/\text{IMERG}$) across intensity bins are displayed for (\textbf{c}) the tropics and (\textbf{d}) the extratropics, where the dotted horizontal line indicates perfect agreement ($1.0$). ERA5 exhibits a persistent overestimation of light rain (<3~mm\,day$^{-1}$) and an underestimation of heavy precipitation extremes across both regions. }
    \label{fig:SI_precip_intensity_ratios}
\end{figure}

\begin{figure}[ht]
    \centering
    \includegraphics[width=\textwidth]{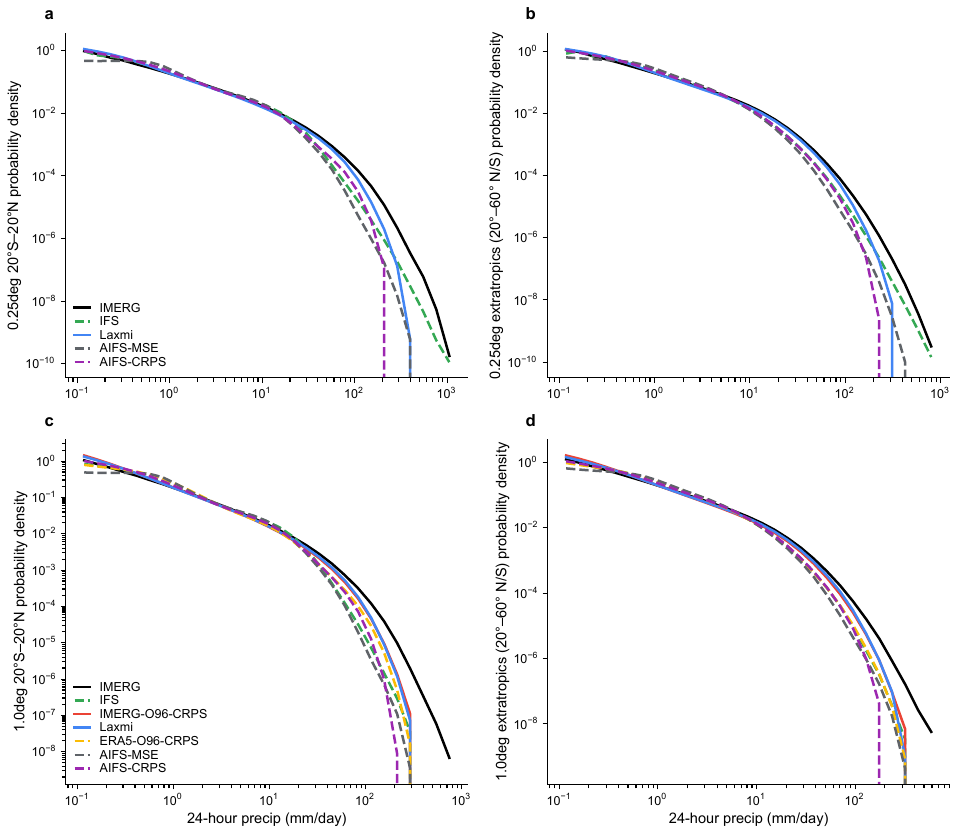}
    \caption{\textbf{Regional precipitation probability density functions.} Probability density distributions of 24-hour accumulated precipitation for (\textbf{a}) the tropics at 0.25$^\circ$, (\textbf{b}) the extratropics at 0.25$^\circ$, (\textbf{c}) the tropics at 1.0$^\circ$, and (\textbf{d}) the extratropics at 1.0$^\circ$. Solid lines distinguish models trained directly on IMERG observations, while dashed lines denote models trained solely on ERA5 reanalysis data or the physical IFS model.}
    \label{fig:figure2_intensity}
\end{figure}

\begin{figure}
    \centering
    \includegraphics[width=\textwidth]{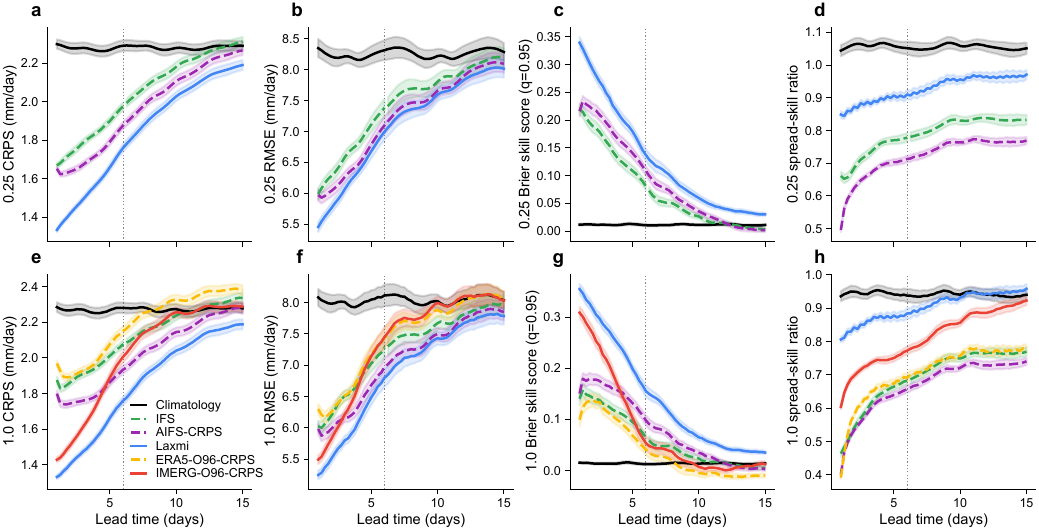}
    \caption{\textbf{Forecasting scores for 24-hour accumulated precipitation evaluated against IMERG.} Panels (\textbf{a}--\textbf{d}) are identical to Figure~2 which evaluate models between $60^\circ\text{S}$ and $60^\circ\text{N}$ for (\textbf{a}) CRPS, (\textbf{b}) unbiased ensemble mean RMSE, (\textbf{c}) Brier skill score, and (\textbf{d}) spread-skill ratio evaluated at $0.25^\circ$ resolution. Panels (\textbf{e}--\textbf{h}) evaluate the models at 1$^\circ$ in order to compare the effect of training resolution and includes ERA5-O96-CRPS and IMERG-O96-CRPS. Solid curves represent models trained to IMERG, including Laxmi and IMERG-O96-CRPS, while dashed curves indicate models trained against ERA5, including AIFS-CRPS and ERA5-O96-CRPS. The IFS is also shown for both resolutions. Each model was initialized 94 times with 50 ensemble members weekly throughout 2024 at 06:00 and 18:00 UTC. After 6 days, denoted by the vertical dotted line, evaluations for IFS change to the 00:00 and 12:00 UTC forecasts. This may give IFS a slight advantage~\cite{Lam2023-tj} and therefore doesn't compromise our conclusion that Laxmi exhibits superior medium-range skill past day 6. Shading for each curve represents a 95\% confidence interval for the average score over the initialization times.}
    \label{fig:figure1_both_resolutions}
\end{figure}

\begin{figure}
    \centering
    \includegraphics[width=\textwidth]{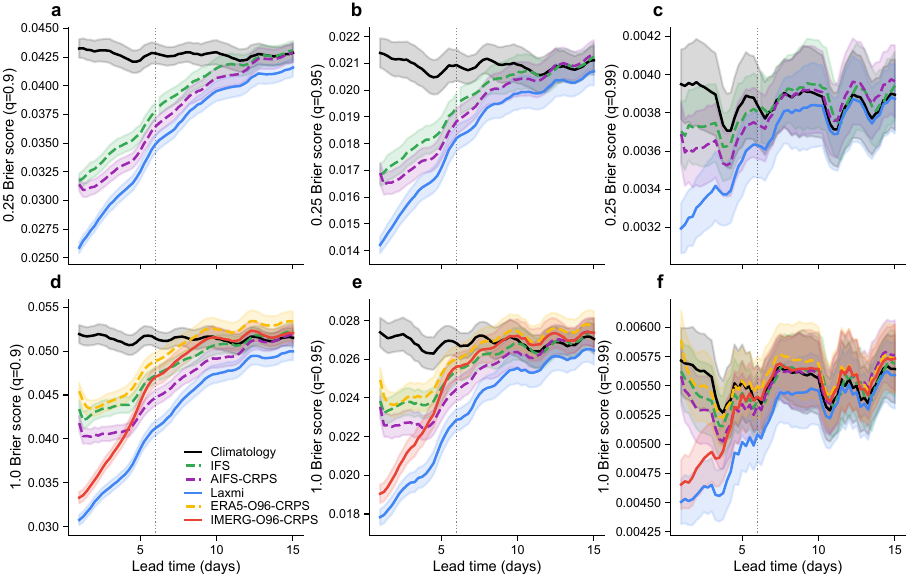}
    \caption{\textbf{Global Brier scores for extreme precipitation percentiles.} Evaluation of Brier scores at the (\textbf{a}, \textbf{d}) 90\textsuperscript{th}, (\textbf{b}, \textbf{e}) 95\textsuperscript{th}, and (\textbf{c}, \textbf{f}) 99\textsuperscript{th} precipitation quantiles against IMERG observations. The models are evaluated over the global domain ($60^\circ\text{S}$--$60^\circ\text{N}$) at both (\textbf{a}--\textbf{c}) 0.25$^\circ$ and (\textbf{d}--\textbf{f}) 1$^\circ$ spatial resolutions. Initialization dates, ensemble size, and models are identical to those described in Figure~2 with the addition of two ${\sim}1^\circ$ models: ERA5-O96-CRPS and IMERG-O96-CRPS.}
    \label{fig:SI_bs_global}
\end{figure}

\begin{figure}[ht]
    \centering
    \includegraphics[width=\textwidth]{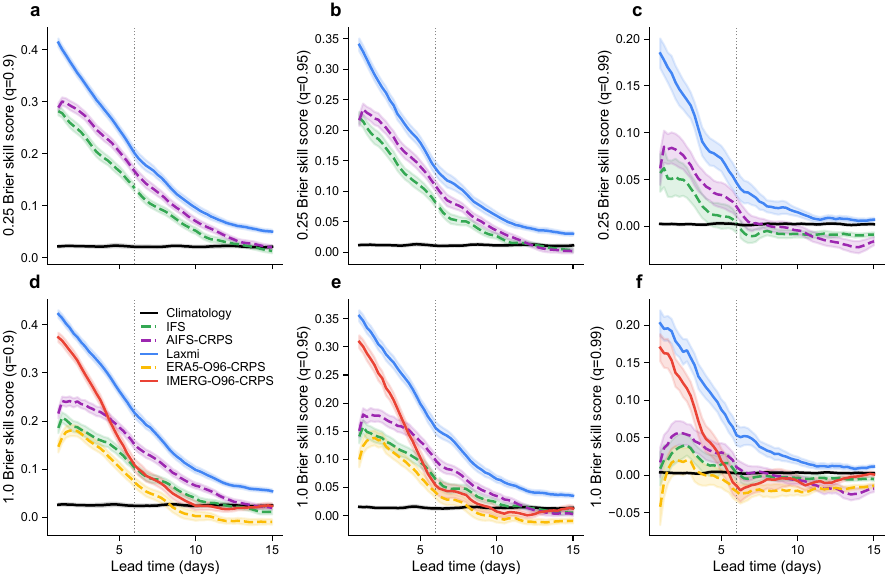}
    \caption{\textbf{Global Brier skill scores for extreme precipitation percentiles.} Evaluation of Brier skill scores at the (\textbf{a}, \textbf{d}) 90\textsuperscript{th}, (\textbf{b}, \textbf{e}) 95\textsuperscript{th}, and (\textbf{c}, \textbf{f}) 99\textsuperscript{th} precipitation quantiles against IMERG observations. The models are evaluated over the global domain ($60^\circ\text{S}$--$60^\circ\text{N}$) at both (\textbf{a}--\textbf{c}) 0.25$^\circ$ and (\textbf{d}--\textbf{f}) 1$^\circ$ spatial resolutions against the probabilistic climatology (see Methods). Initialization dates, ensemble size, and models are identical to those described in Figure~2 with the addition of two ${\sim}1^\circ$ models: ERA5-O96-CRPS and IMERG-O96-CRPS.}
    \label{fig:SI_bss_global}
\end{figure}

\begin{figure}
    \centering
    \includegraphics[width=\textwidth]{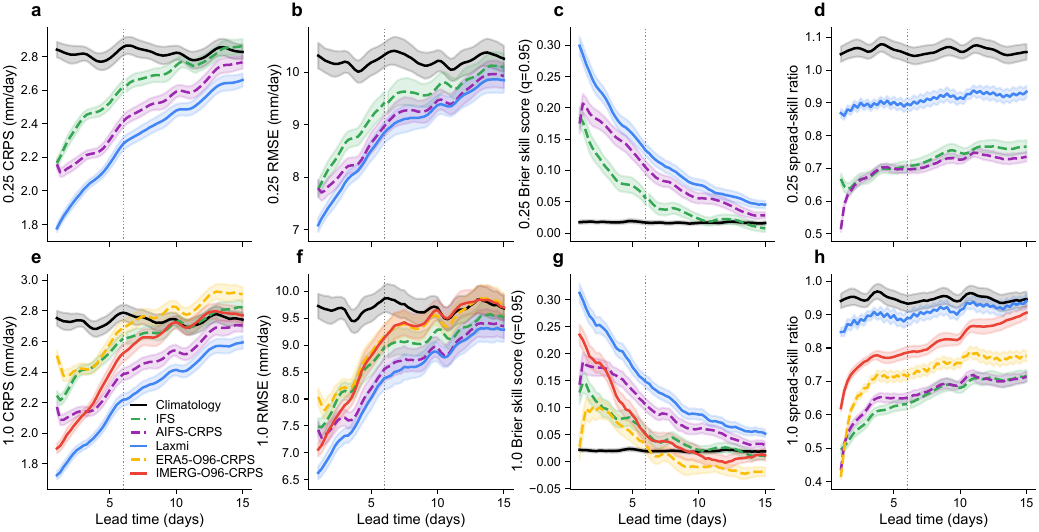}
    \caption{\textbf{Forecasting scores for 24-hour accumulated precipitation evaluated against IMERG in the tropics.} Same as Figure~2, but with evaluation restricted to the tropics (20$^\circ$S--20$^\circ$N) for (\textbf{a}) CRPS, (\textbf{b}) unbiased ensemble mean RMSE, (\textbf{c}) Brier skill score, and (\textbf{d}) spread-skill ratio at 0.25$^\circ$ resolution. (\textbf{e}--\textbf{h}) The metrics are also evaluated at 1$^\circ$ resolution and include two additional models trained at lower resolution: ERA5-O96-CRPS and IMERG-O96-CRPS. Initialization dates, ensemble size, and models are identical to those described in Figure~2.}
    \label{fig:SI_fig1_tropics}
\end{figure}

\begin{figure}
    \centering
    \includegraphics[width=\textwidth]{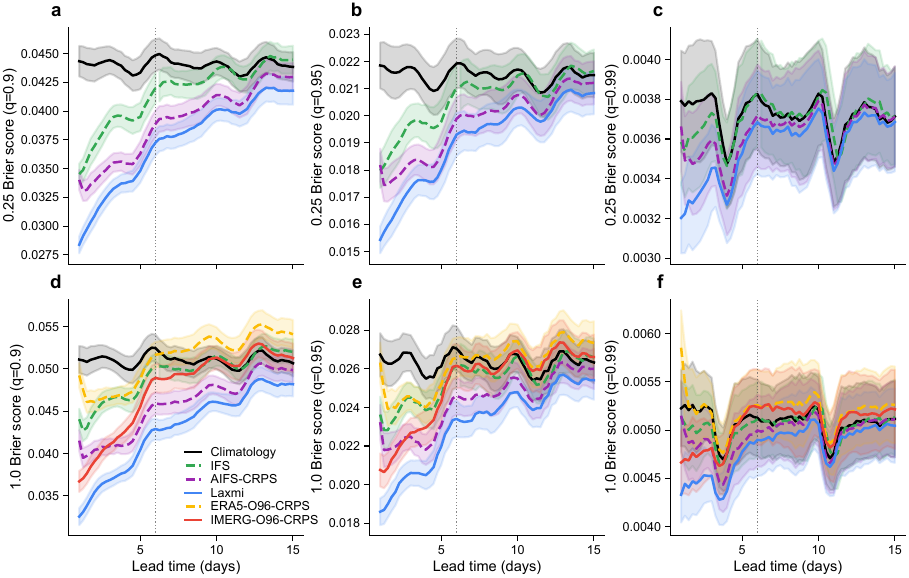}
    \caption{\textbf{Tropical Brier scores for extreme precipitation percentiles.} Evaluation of Brier scores at the (\textbf{a}, \textbf{d}) 90\textsuperscript{th}, (\textbf{b}, \textbf{e}) 95\textsuperscript{th}, and (\textbf{c}, \textbf{f}) 99\textsuperscript{th} precipitation quantiles against IMERG observations. The models are evaluated in the tropics ($20^\circ\text{S}$--$20^\circ\text{N}$) at both (\textbf{a}--\textbf{c}) 0.25$^\circ$ and (\textbf{d}--\textbf{f}) 1$^\circ$ spatial resolutions. Initialization dates, ensemble size, and models are identical to those described in Figure~2 with the addition of two ${\sim}1^\circ$ models: ERA5-O96-CRPS and IMERG-O96-CRPS.}
    \label{fig:SI_bs_tropics}
\end{figure}

\begin{figure}[ht]
    \centering
    \includegraphics[width=\textwidth]{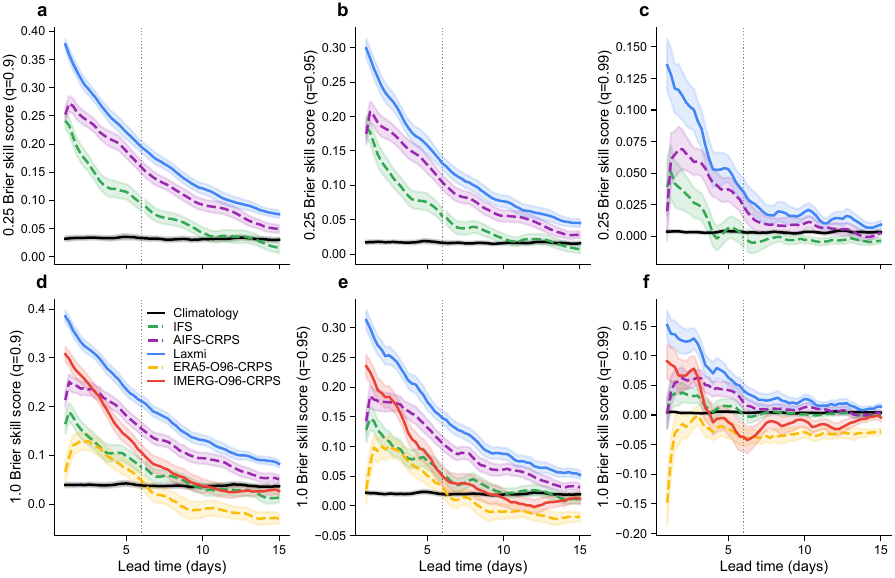}
    \caption{\textbf{Tropical Brier skill scores for extreme precipitation percentiles.} Evaluation of Brier skill scores at the (\textbf{a}, \textbf{d}) 90\textsuperscript{th}, (\textbf{b}, \textbf{e}) 95\textsuperscript{th}, and (\textbf{c}, \textbf{f}) 99\textsuperscript{th} precipitation quantiles against IMERG observations. The models are evaluated in the tropics ($20^\circ\text{S}$--$20^\circ\text{N}$) at both (\textbf{a}--\textbf{c}) 0.25$^\circ$ and (\textbf{d}--\textbf{f}) 1$^\circ$ spatial resolutions against the probabilistic climatology~(see Methods). Initialization dates, ensemble size, and models are identical to those described in Figure~2 with the addition of two ${\sim}1^\circ$ models: ERA5-O96-CRPS and IMERG-O96-CRPS.}
    \label{fig:SI_bss_tropics}
\end{figure}

\begin{figure}[ht]
    \centering
    \includegraphics[width=\textwidth]{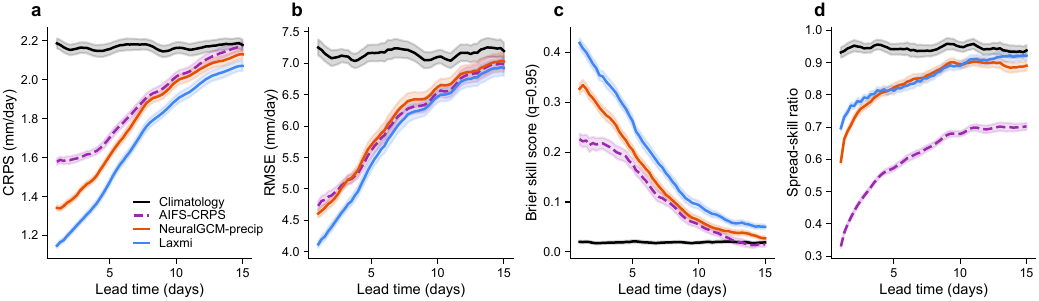}
    \caption{\textbf{Comparison of AIFS-CRPS, NeuralGCM-precip, and Laxmi at $\sim 2.8^\circ$ resolution.} We compare our model to another AIWP model, NeuralGCM-precip~\cite{Yuval2026-dd}, that was trained directly on IMERG precipitation observations. Metrics shown here are the same as Figure~2 with panels (\textbf{a}) CRPS, (\textbf{b}) unbiased ensemble mean RMSE, (\textbf{c}) Brier skill score, and (\textbf{d}) spread skill ratio. Model initializations include 104 initializations at 00:00 and 12:00 UTC for 2024.}
    \label{fig:SI_metrics_global_ngcm}
\end{figure}

\begin{figure}[ht]
    \centering
    \includegraphics[width=\textwidth]{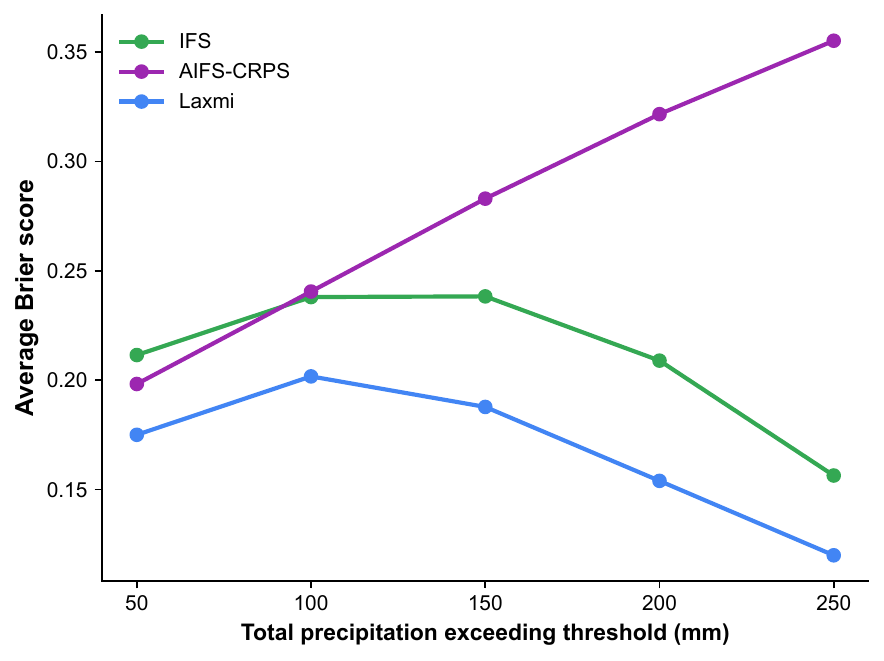}
    \caption{\textbf{Average Brier scores for the IFS ensemble, AIFS-CRPS, and Laxmi as a function of precipitation threshold for all 10 landfalling tropical systems in India between February 8\textsuperscript{th}, 2024 and September 30\textsuperscript{th}, 2025.}}
    \label{fig:SI_average_brier_score_NI}
\end{figure}

\begin{figure}[ht]
    \centering
    \includegraphics[width=\textwidth]{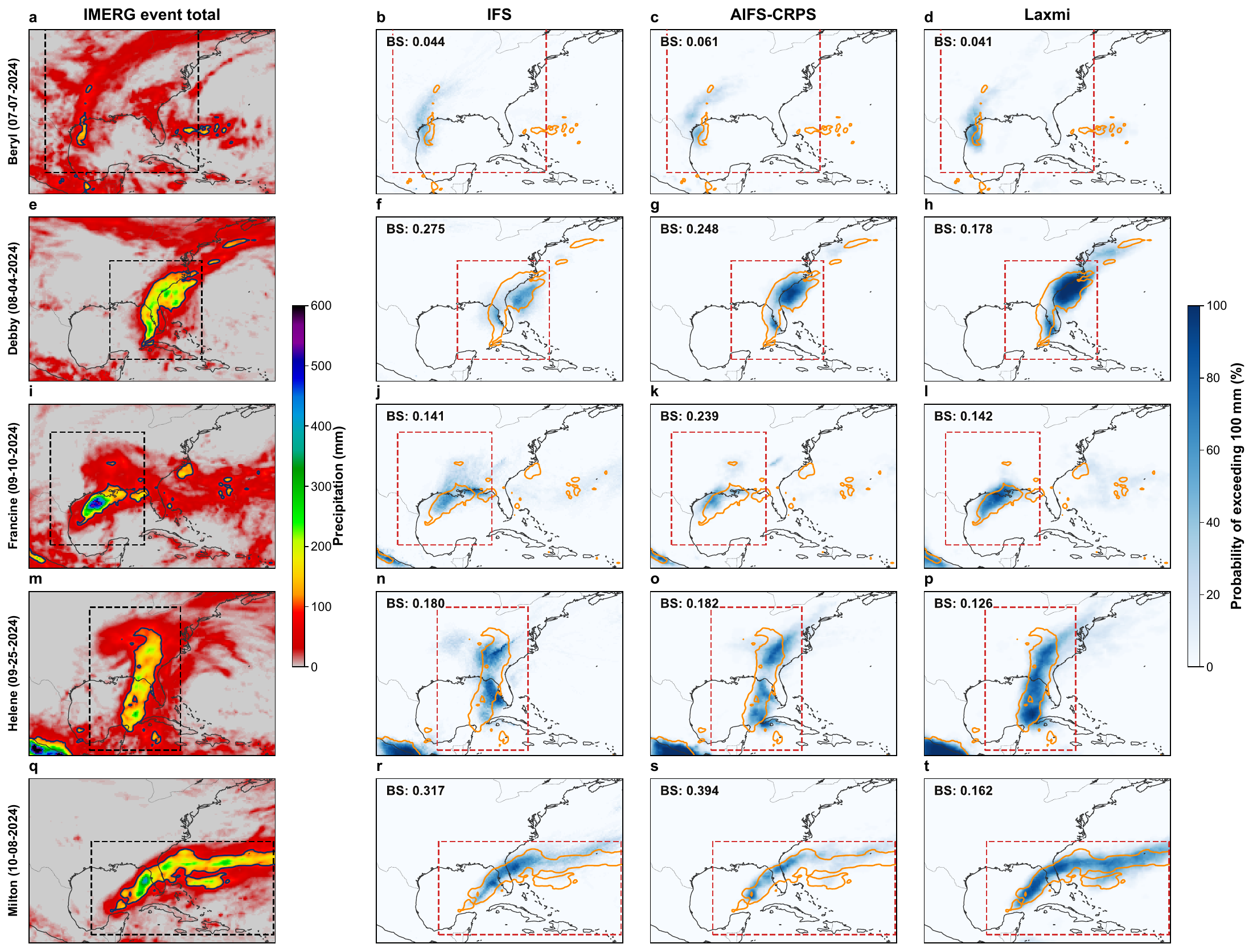}
    \caption{\textbf{Observed and forecasted extreme precipitation for the five landfalling tropical cyclones in the U.S. in 2024 and in 2025 before September 30\textsuperscript{th}, 2025. } Rows display event-integrated accumulation (spanning 24 hours before to 72 hours after landfall) for hurricanes Beryl (\textbf{a}--\textbf{d}), Debby (\textbf{e}--\textbf{h}), Francine (\textbf{i}--\textbf{l}), Helene (\textbf{m}--\textbf{p}), and Milton (\textbf{q}--\textbf{t}). Forecasts were initialized at least 72 hours before the landfall time listed in IBTrACS~\cite{R2019-cs} at 06:00 or 18:00 UTC. Columns show observed IMERG accumulated precipitation (\textbf{a}, \textbf{e}, \textbf{i}, \textbf{m}, \textbf{q}) alongside model probability of exceedance (PoE) over 50-member ensembles at the 100~mm threshold for IFS (\textbf{b}, \textbf{f}, \textbf{j}, \textbf{n}, \textbf{r}), operational AIFS-CRPS (\textbf{c}, \textbf{g}, \textbf{k}, \textbf{o}, \textbf{s}), and Laxmi (\textbf{d}, \textbf{h}, \textbf{l}, \textbf{p}, \textbf{t}). Solid contours overlaid on forecast panels indicate the observed region where IMERG accumulated precipitation exceeded 100~mm. Annotated values in the upper left of each forecast panel report the event-level Brier score evaluated at the 100~mm accumulation threshold defined across the active region (see Methods), which is indicated by the dashed box.}
    \label{fig:SI_case_study_us_hurricanes}
\end{figure}

\begin{figure}[ht]
    \centering
    \includegraphics[width=\textwidth]{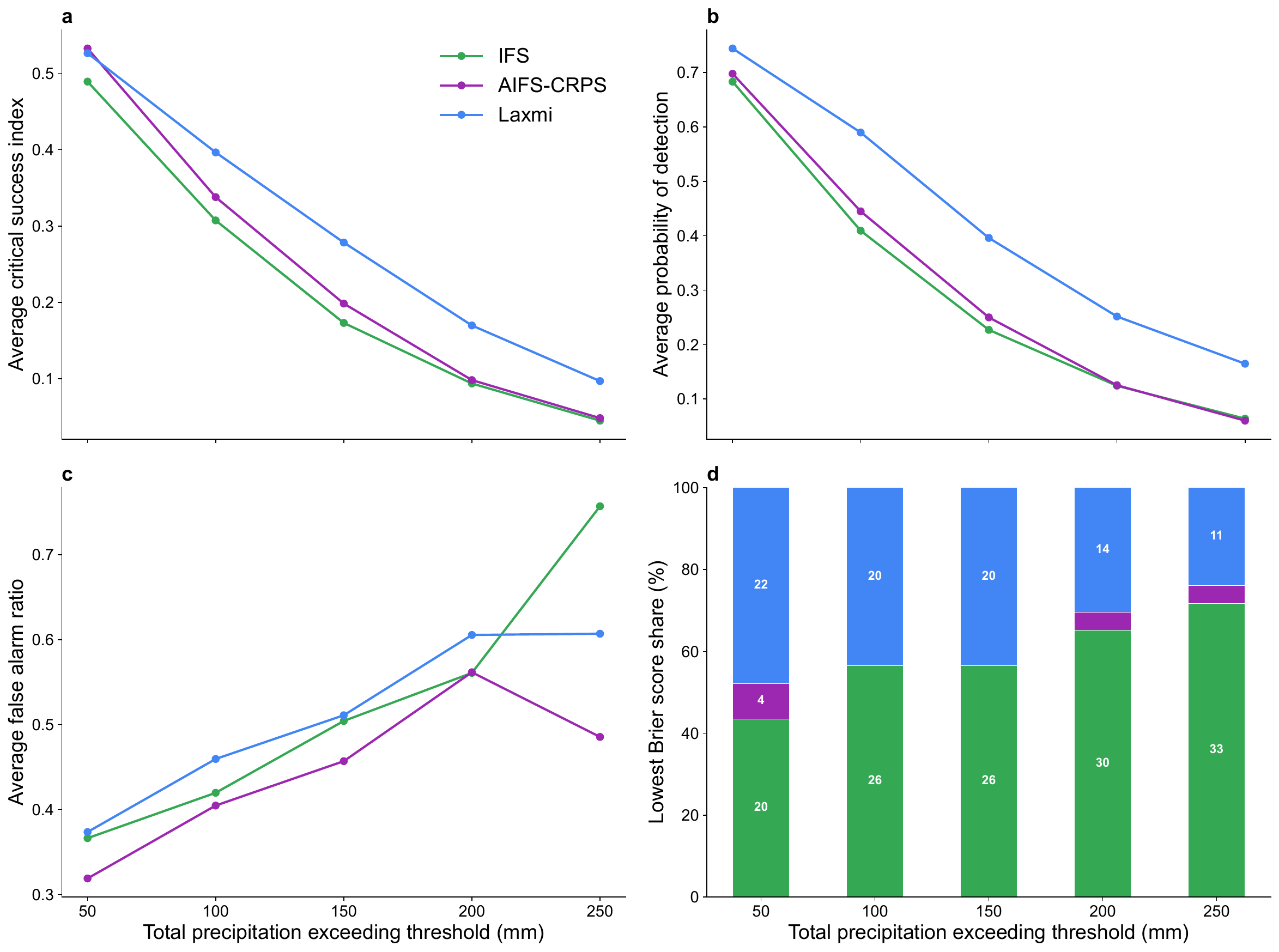}
    \caption{\textbf{Evaluation of decision-oriented forecast skill across 46 globally landfalling tropical cyclones.} Comparison of IFS, AIFS-CRPS, and Laxmi for 46 tropical cyclones that made landfall globally between 8 February 2024 and 30 September 2025~\cite{Knapp2010-sx, R2019-cs}, evaluated across five precipitation thresholds (50, 100, 150, 200, and 250~mm\,day$^{-1}$). Across all 46 events, we compute (\textbf{a}) Critical Success Index, (\textbf{b}) Probability of Detection, (\textbf{c}) False Alarm Rate at each of the thresholds and also include (\textbf{d}) the best forecast by Brier Score at the 100~mm threshold over the active region (see Methods). The active region is defined as the area within 500~km of the IBTrACS best track estimate, and further restricted to the union of the regions where any of the three models produce a non-zero probability of exceeding the precipitation threshold. Categorical metrics (CSI, POD, and FAR) are computed using a 30\% probability warning threshold (i.e., a grid cell is classified as a warning if at least 15 of 50 ensemble members exceed the precipitation target).}
    \label{fig:SI_metrics_us_hurricanes}
\end{figure}

\clearpage
\bibliography{bibliography}